# Fluid-structure interaction study of spider's hair flow-sensing system

Roberto Guarino[a], Gabriele Greco[a,b], Barbara Mazzolai[b], Nicola M. Pugno[a,c,d,]*

[a]Laboratory of Bio-Inspired & Graphene Nanomechanics, Department of Civil, Environmental and Mechanical Engineering, University of Trento, Via Mesiano 77, 38123 Trento, Italy
[b]Center for Micro-BioRobotics@SSSA, Istituto Italiano di Tecnologia, Viale Rinaldo Piaggio 34, 56025 Pontedera, Italy
[c]Ket Lab, Edoardo Amaldi Foundation, Italian Space Agency, Via del Politecnico snc, 00133 Rome, Italy
[b]School of Engineering and Materials Science, Queen Mary University of London, Mile End Road, London E1 4NS, United Kingdom

**Abstract**

In the present work we study the spider's hair flow-sensing system by using fluid-structure interaction (FSI) numerical simulations. We observe experimentally the morphology of *Theraphosa stirmi*'s hairs and characterize their mechanical properties through nanotensile tests. We then use the obtained information as input for the computational model. We study the effect of a varying air velocity and a varying hair spacing on the mechanical stresses and displacements. Our results can be of interest for the design of novel bio-inspired systems and structures for smart sensors and robotics.

*Keywords:* sensing system; smart sensor; spider hair; numerical simulation; fluid-structure interaction

## 1. Introduction

The ability of spiders in colonizing many habitats is also due to the fact that they interact efficiently with other animals for catching preys, escaping from predators and detecting potential partners also during the night. Spiders present a very complex and advanced sensing system based on tactile (or mechanical) [1-4], chemical [5,6] and air flow receptors [7-9] such as the hairs. These latter structures interact with the air velocity field and provide the spider with information regarding the surrounding environment [10,11], representing one of the most sensitive biosensors in Nature [12]. Since their extraordinary sensibility and efficiency, the interest in designing bio-inspired sensing systems and structures for soft robotics and high-tech applications has increased in the last decade [13-15].

Several numerical simulation works are available in the literature, but they are mainly related, for instance, to the engineering design and analysis of bio-inspired micro electro-mechanical systems (MEMS) (see, e.g., Ref. [16]). Numerical studies on the biological structures, in fact, are still poorly addressed and further investigations are needed in order to characterize the performance of spiders' hairs.

---

* Corresponding author. Tel.: +39-0461-282525; fax: +39-0461-282599.
  E-mail address: nicola.pugno@unitn.it

In this work, we employ fluid-structure interaction (FSI) simulations to study the behavior of spider's hairs in air. We investigate the effect of a varying flow velocity on the sensing capabilities, quantified in terms of displacements and von Mises stresses; and of a varying spacing of the hairs. The morphology of the hairs and their elastic properties are obtained experimentally, through Scanning Electron Microscopy (SEM), optical microscopy and nanotensile tests, and are used as inputs in the numerical simulations.

## 2. Materials and methods

*2.1. Sample preparation*

The hairs are obtained from an exoskeleton of *Theraphosa stirmi* kept under controlled feeding and environmental conditions. The tested samples are prepared following the same procedure reported by Blackledge *et al.* [17]. We stick the hair samples on a paper frame provided with a square window of 5 mm side. The hair sample is fixed on the paper frame with a double-sided tape coupled with a glue.

*2.2. Morphology and mechanical properties*

For the morphology characterization by SEM, we use a Zeiss Supra 40 (Carl Zeiss AG, Germany) at 2.30 kV. The metallization is made by using a sputtering machine Q150T (Quorum Technologies Ltd, UK) and the sputtering mode is Pt/Pd 80:20 for 5 minutes. In addition, we employ optical microscopy (Carl Zeiss AG, Germany) at 10x magnification for the extraction of an average value of the hair diameter.

For the mechanical characterization, we use a T150 UTM nanotensile machine (Agilent Technologies Inc., USA) with a 500 mN load cell. The displacement speed is 10 µm s$^{-1}$ with the frequency load at 20 Hz. The declared sensitivity of the machine is 10 nN for the load and 1 Å for the displacement in the dynamic configuration.

*2.3. Numerical simulations*

The FSI simulations are carried out in COMSOL Multiphysics® [18]. We employ two-way coupling in order to have a comprehensive understanding of the interaction between fluid flow and mechanical displacements [19].

A single spider's hair is considered and it is approximated with a truncated cone of height $L = 1000$ µm, base diameter $D_b = 30$ µm and tip diameter $D_t = 15$ µm. The computational domain for the fluid flow is composed of a 3D box of width 100 µm and height 1200 µm, with inlet and outlet positioned 1000 µm before and after the hair, respectively. Moreover, a chitin of thickness $h = 100$ µm is considered below the hair. We assign linear elastic material properties, using a Young's modulus $E = 600$ MPa (i.e. in the order of the measured value), a Poisson's ratio $v = 0.25$ and a density $\rho_h = 1425$ kg m$^{-3}$ (taken close to the respective values for chitin [20]). For the air, we consider its properties at 300 K, i.e. density $\rho_a = 1.177$ kg m$^{-3}$ and dynamic viscosity $\mu_a = 1.85 \cdot 10^{-5}$ kg m$^{-1}$ s$^{-1}$ [21]. A constant fluid velocity is imposed at the inlet and the domain is delimited by walls with full-slip boundary conditions. The whole domain is discretized with about $4.5 \cdot 10^5$ tetrahedral elements.

The inlet velocity is chosen from 0.1 to 1 m s$^{-1}$, which is in the range of the velocity field measured around flying insects [10]. The corresponding maximum von Mises stress $\sigma_{max}$ and maximum displacement $\delta_{max}$ is measured. An additional analysis is carried out on the hair spacing, considering two hairs on the same transversal coordinate, but positioned from 50 to 1000 µm from each other. In this case, the differences in $\sigma_{max}$ and $\delta_{max}$ are monitored, for a constant mean inlet velocity $u_{avg} = 1$ m s$^{-1}$.

## 3. Results

*3.1. Morphology and mechanical properties*

Fig. 1 shows the SEM images of different hair typologies found on the spiders' exoskeleton. We can observe a typical length from a few hundreds of µm to 1-2 mm and a variable diameter, with a tip size usually in the range 1-20 µm. The typical hair spacing is a few mm, i.e. in the order of the hair length. Usually, shorter hairs not involved in flow-sensing are also present on the spider's cuticle and are not shown here, for brevity.

The stress-strain curves extracted from the nanotensile test on different hairs are shown in Fig. A1. The mechanical properties of the hairs are intrinsically variable, because of their biological nature, and we have measured an initial Young's modulus in the range 500-800 MPa using linear regression, with a fracture strain in the order of 0.5 mm mm$^{-1}$. In Fig. A2, we show a *Theraphosa stirmi* of the collection and an optical microscopy investigation. The average diameter of the hairs is around 30 µm.

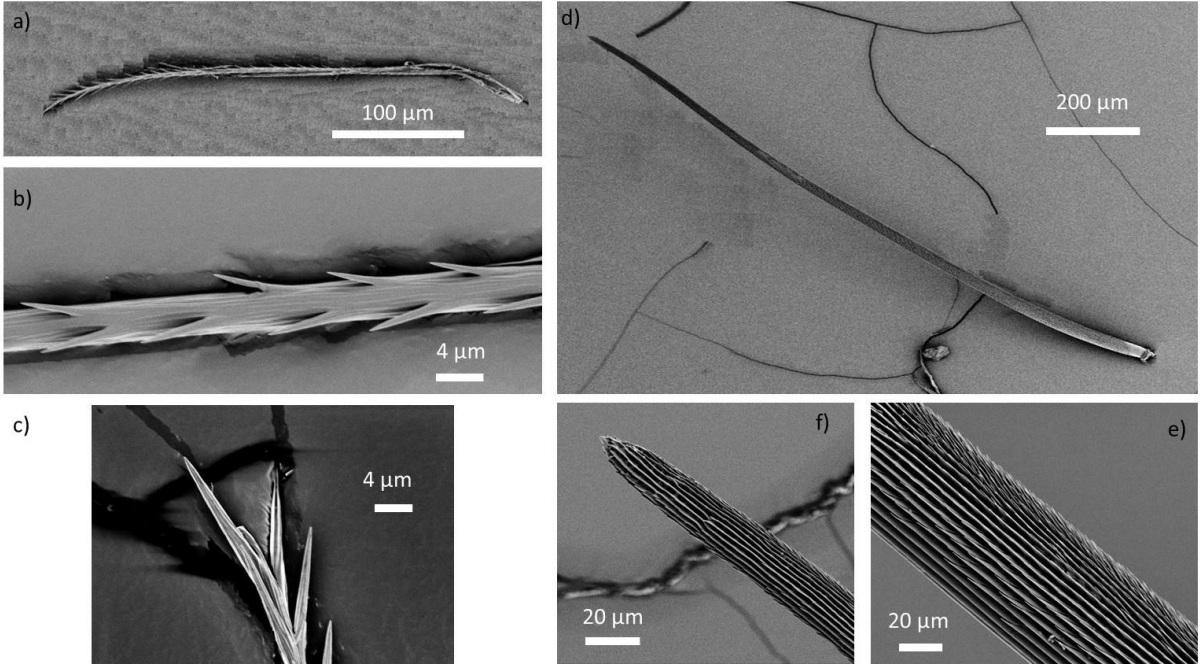

Fig. 1. SEM morphology of different hairs found on the spider's exoskeleton.

## 3.2. Effect of a varying air flow

Fig. 2 shows the values of $\sigma_{max}$ and $\delta_{max}$ as function of the mean inlet velocity, as well as the von Mises stress distribution together with the qualitative flow streamlines. We can observe that both the stress and the displacement scale almost linearly with $u_{avg}$.

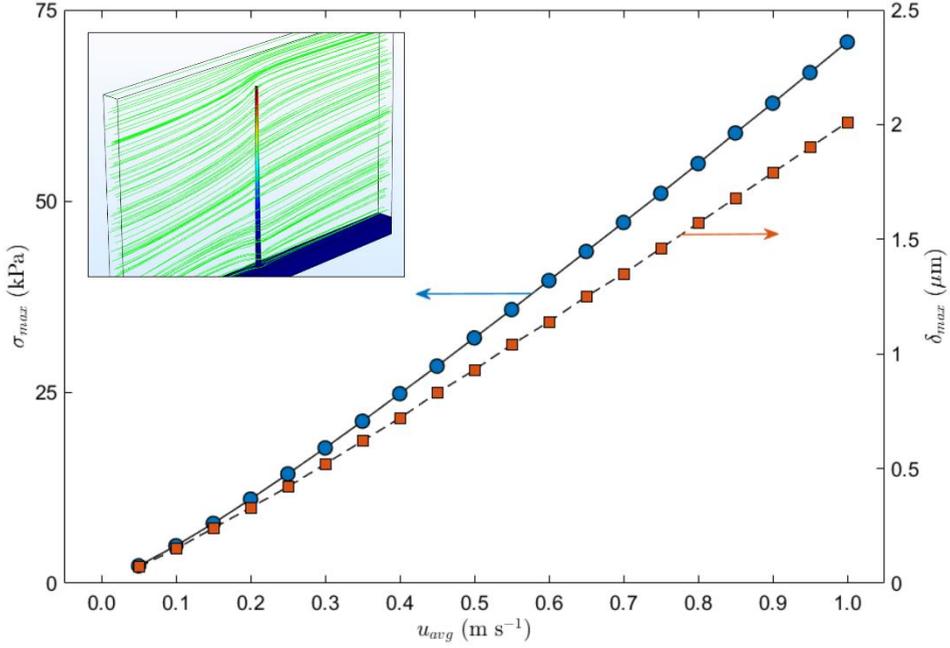

Fig. 2. Maximum von Mises stress and maximum hair displacement as function of the mean inlet velocity. Inset: qualitative visualization of the displacement field of the structure and of the flow streamlines.

A rough estimation of the mechanical load can be obtained by considering a beam, here approximated as a cylinder clamped at one end, with average diameter $D_{avg} = (D_b + D_t)/2$. We can assume that the fluid flow exerts a constant force per unit length, which, denoting by $C_D$ the dimensionless drag coefficient, is given by:

$$f_D = \frac{1}{2} \rho_a C_D D_{avg} u_{avg}^2 \tag{1}$$

Neglecting the shear deformation, the displacement at the tip of the hair is given by:

$$\delta_{max} \approx \frac{f_D L^4}{8EI} = \frac{4\rho_a C_D L^4}{\pi E D_{avg}^3} u_{avg}^2 \tag{2}$$

where we have inserted the expression of the maximum bending moment, equal to $f_D L^2/2$, and the moment of inertia of the section, i.e. $I = \pi D_{avg}^4 / 64$.

For a cylinder immersed in a laminar flow, for low Reynolds numbers Re and considering constant fluid properties, the drag coefficient depends on the flow velocity roughly as [22]:

$$C_D = C_D(u_{avg}) \sim \text{Re}^{-1} \sim u_{avg}^{-1} \tag{3}$$

Therefore, inserting Eq. (3) into Eq. (2), we get the observed linear relationship between maximum hair displacement and mean inlet velocity, i.e. $\delta_{max} \sim u_{avg}$. An analogous reasoning can be made on the von Mises stress, except for the fact that its expression is more complicated because it must include, together with the bending stress, also a shear term and the stresses deriving from the substrate. For simplicity, considering only the maximum bending stress $\sigma_{b,\max}$:

$$\sigma_{b,\max} \approx \frac{f_D L^2}{2I} \frac{D_{avg}}{2} = \frac{8\rho_a C_D L^2}{\pi D_{avg}^2} u_{avg}^2 \qquad (4)$$

and again, from Eq. (3), we get a law of the type $\sigma_{b,\max} \sim u_{avg}$ as shown in Fig. 2. Since Eq. (2) and Eq. (4) are obtained considering a rigid substrate, we can observe that they underestimate the maximum displacement at the hair tip and overestimate the maximum stress at the hair base, respectively.

We can fit the values of the maximum displacement with a law of the type $\delta_{\max} = K_\delta u_{avg}$, obtaining $K_\delta \approx 1.943 \cdot 10^{-6}$ s ($R^2$-value 0.9944). Consequently, the numerically-derived expression of the drag coefficient is:

$$C_D \approx 12.7 \, \mathrm{Re}^{-1} \qquad (5a)$$

Considering the expression of the maximum bending stress in Eq. (4), instead, we can employ a law of the type $\sigma_{b,\max} = K_\sigma u_{avg}$ for fitting, obtaining $K_\sigma \approx 67840$ kg m$^{-2}$ s$^{-1}$ ($R^2$-value 0.9917). In this case, the drag coefficient is:

$$C_D \approx 16.4 \, \mathrm{Re}^{-1} \qquad (5b)$$

Both expression of $C_D$ are in the same order of magnitude of the data available in the literature, e.g. for perfect spheres or cylinders [22]. Furthermore, we can observe that Eq. (5a) can be considered as a lower bound, while Eq. (5b) as an upper bound, because of the rigid substrate assumption described above. This result represents a good estimation of the aerodynamic performance of the spider's hair at low Re, if considering the introduced approximation (i.e. cylindrical instead of conical geometry and rigid constraint) and the presence of the hair tip (i.e. finite-length body). Note that a more general power law can be employed for fitting, e.g. $\delta_{\max} = K_1 u_{avg}^{K_2}$ for the displacement, and we get $K_2 \approx 1.125$ ($R^2$-value 0.9999) thus the linear law assumed above can be considered sufficiently precise.

Usually sensors are based on linear responses, since only one parameter is necessary to correlate the quantity to measure to an electric signal, and we observe that the spider's sensing system present a similar behavior. Thus, this effect can be particularly interesting for the design of new sensors and MEMS for fluid flow measurements.

*3.3. Effect of a varying hair spacing*

The effect of the hair spacing, here denoted with $\Delta s$, is quantified by monitoring the response of the second hair with respect to the first one. Specifically, we measure the variation of the maximum displacement at the hair tip and of the maximum von Mises stress at the hair base. As shown in Fig. 3, for small values of $\Delta s$ the effect of the aerodynamic wake behind the first hair is to reduce the stresses and displacements to which the second hair is subjected. Therefore, for small values of the spacing there is a significant variation of $\sigma_{\max}$ and $\delta_{\max}$, while it becomes negligible (or, at least, acceptable for the sensing system, being below 10%) when $\Delta s$ is in the order of half-length of the hair (i.e. around 500 µm for the two 1000 µm hairs considered here). Note that this result has been obtained considering the maximum flow velocity, i.e. $u_{avg} = 1$ m s$^{-1}$. For lower values of $u_{avg}$, the variations $\Delta\sigma_{\max}$ and $\Delta\delta_{\max}$ are expected to be lower due to a shorter aerodynamic wake, thus the sensing system is effective also for lower values of the hair spacing.

This result suggests that the relative position of flow-sensing hairs is probably dictated by the mutual interference. Therefore, we have observed that there is an optimal configuration to maximize the spider's sensing capabilities: this happens when there is the maximum density of hairs (i.e. number of hairs per unit length of the substrate) capable of sensing a certain level of stress. This optimal density of hairs is compatible to what is observed experimentally, as shown in Fig. A3.

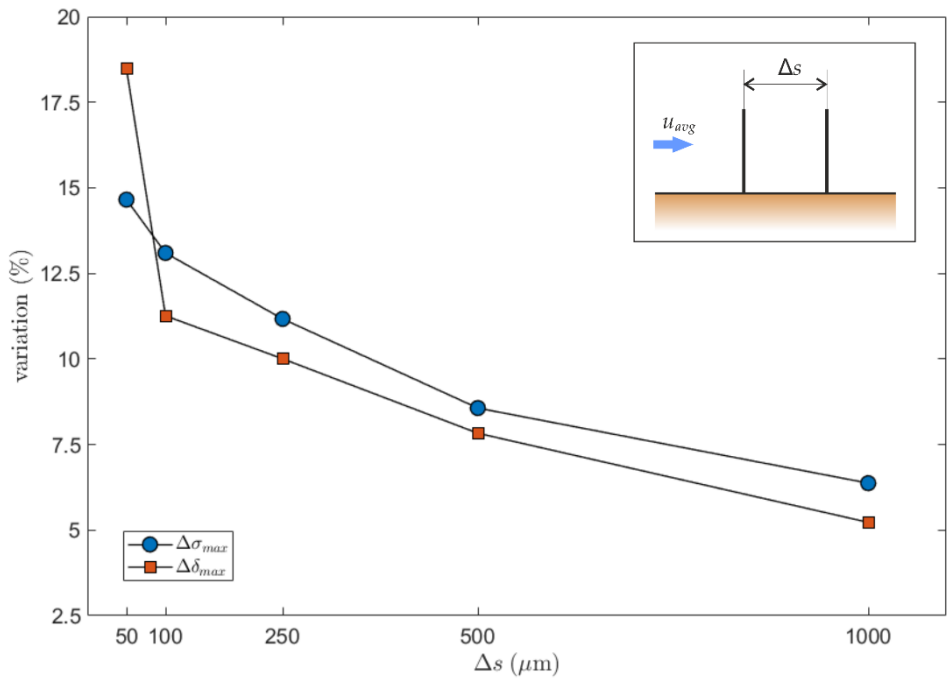

Fig. 3. Difference on the maximum von Mises stress and the maximum displacement between the first and the second hair as function of the hair spacing. Inset: schematic view of the considered geometry.

## 4. Conclusions

We have used FSI numerical simulations to investigate the behavior of *Theraphosa stirmi*'s hairs under a fluid flow. We have extracted the morphology and the elastic properties of the hairs from experimental tests. The simulation study has allowed to compute the mechanical response of the hair under a varying flow velocity, suggesting a linear relationship between maximum displacement and mean inlet velocity, as well as between maximum von Mises stress and $u_{avg}$. In addition, we have considered the spacing between two hairs, showing that when the spacing is low the wake of the first hair affects the sensing capability of the second one. Therefore, this suggests that the optimal hair spacing in spiders might derive from fluid dynamic reasons and from the need of maximizing the sensing capabilities. These results can be of interest for the optimal design of novel sensors and MEMS for fluid flow measurements.

## Acknowledgements

RG is supported by Bonfiglioli Riduttori SpA. NMP is supported by the European Commission under the Graphene Flagship Core 2 No. 785219 (WP14 "Polymer Composites") and FET Proactive "Neurofibres" grant No. 732344.

**Appendix A. Mechanical and optical characterization**

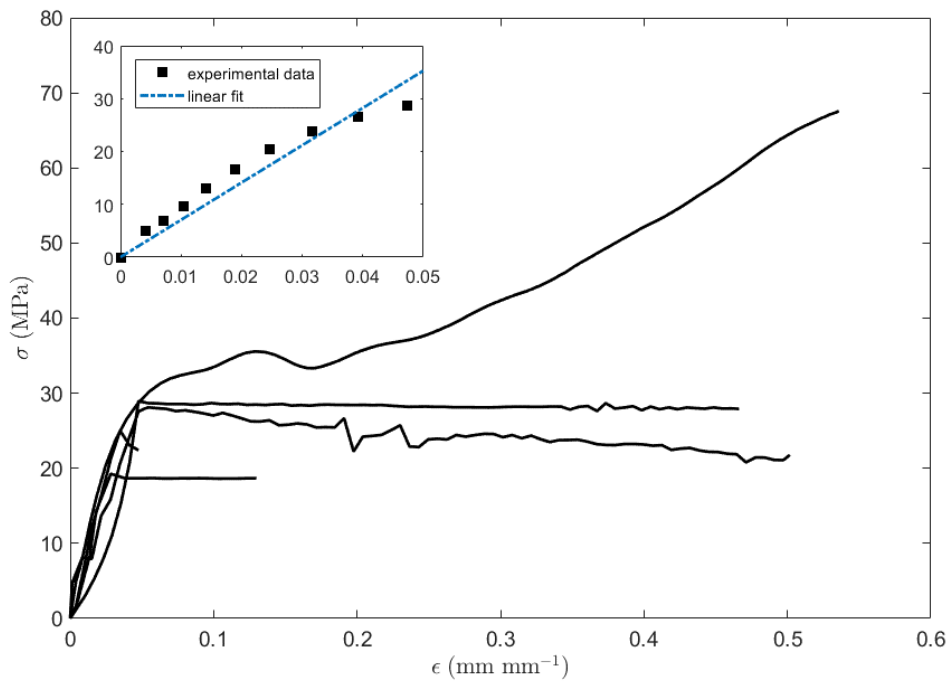

Fig. A1. Example stress-strain curves of different *Theraphosa stirmi*'s hairs obtained through nanotensile tests. Inset: example linear fit on the first part of a curve, with extracted Young's modulus $E \approx 705.8$ MPa ($R^2$-value = 0.921).

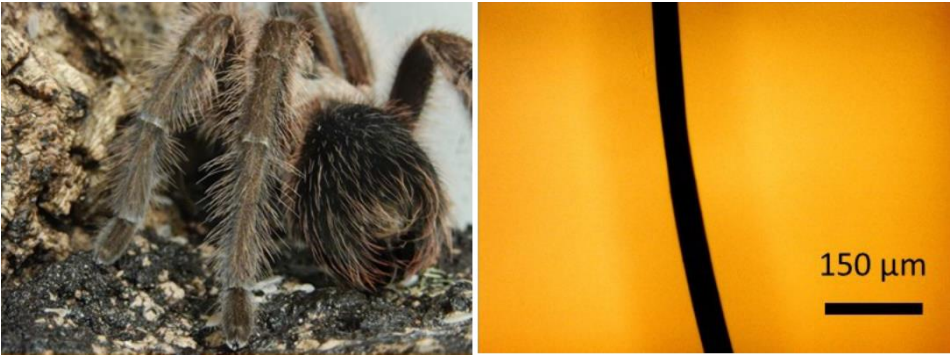

Fig. A2. Left: a picture of a *Theraphosa stirmi* of the collection. Right: optical microscope image of a hair.

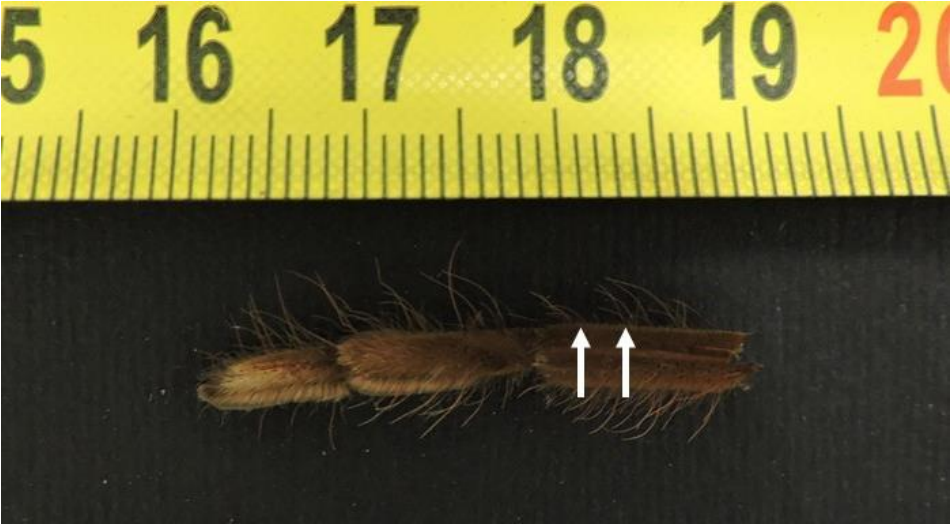

Fig. A3. Picture of a *Theraphosa stirmi*'s leg with flow-sensing hairs and typical spacing in the order of the half-length of a hair (white arrows).